\documentclass[submission,copyright,creativecommons]{eptcs}
\providecommand{\event}{F-IDE 2015} 

\usepackage{etex}
\usepackage{amssymb}
\usepackage{url}

\usepackage[numbered]{bookmark}
\usepackage{layouts}
\usepackage[usenames,dvipsnames,svgnames]{xcolor}
\usepackage[labelfont=bf,labelsep=period]{caption}
\usepackage{environ} 
\usepackage{wrapfig}
\usepackage{amsmath}
\usepackage{pbox}
\usepackage{paralist} 
\usepackage{sidecap}
\usepackage{tikz}
\usetikzlibrary{positioning}
\usepackage{comment}
\usepackage{multirow}
\usepackage{graphicx}
\usepackage{caption}
\usepackage{subcaption} 
\usepackage{bm}
\usepackage{wrapfig}
\usepackage{verbatim}
\usepackage[disable]{todonotes} 

\newcommand*{\missingreference}[1]{\colorbox{black!15!white}{\scriptsize xxx #1}}
\newcommand*{\missingcitation}[1]{\colorbox{black!15!white}{\scriptsize xxx #1}}
\makeatletter
\def\@setref#1#2#3{%
  \ifx#1\relax
   \protect\G@refundefinedtrue
   \nfss@text{\reset@font\missingreference{#3}}
   \@latex@warning{Reference `#3' on page \thepage \space
             undefined}%
  \else
   \expandafter#2#1\null
  \fi}
\def\@citex[#1]#2{\leavevmode
  \let\@citea\@empty
  \@cite{\@for\@citeb:=#2\do
    {\@citea\def\@citea{,\penalty\@m\ }%
     \edef\@citeb{\expandafter\@firstofone\@citeb\@empty}%
     \if@filesw\immediate\write\@auxout{\string\citation{\@citeb}}\fi
     \@ifundefined{b@\@citeb}{\hbox{\reset@font\missingcitation{#2}}
       \G@refundefinedtrue
       \@latex@warning
         {Citation `\@citeb' on page \thepage \space undefined}}%
       {\@cite@ofmt{\csname b@\@citeb\endcsname}}}}{#1}}
\makeatother

\global\long\def\sp{}

\makeatletter
\DeclareFontFamily{OMX}{MnSymbolE}{}
\DeclareSymbolFont{MnLargeSymbols}{OMX}{MnSymbolE}{m}{n}
\SetSymbolFont{MnLargeSymbols}{bold}{OMX}{MnSymbolE}{b}{n}
\DeclareFontShape{OMX}{MnSymbolE}{m}{n}{
    <-6>  MnSymbolE5
   <6-7>  MnSymbolE6
   <7-8>  MnSymbolE7
   <8-9>  MnSymbolE8
   <9-10> MnSymbolE9
  <10-12> MnSymbolE10
  <12->   MnSymbolE12
}{}
\DeclareFontShape{OMX}{MnSymbolE}{b}{n}{
    <-6>  MnSymbolE-Bold5
   <6-7>  MnSymbolE-Bold6
   <7-8>  MnSymbolE-Bold7
   <8-9>  MnSymbolE-Bold8
   <9-10> MnSymbolE-Bold9
  <10-12> MnSymbolE-Bold10
  <12->   MnSymbolE-Bold12
}{}

\let\llangle\@undefined
\let\rrangle\@undefined
\DeclareMathDelimiter{\llangle}{\mathopen}%
                     {MnLargeSymbols}{'164}{MnLargeSymbols}{'164}
\DeclareMathDelimiter{\rrangle}{\mathclose}%
                     {MnLargeSymbols}{'171}{MnLargeSymbols}{'171}
\makeatother

\newcommand{\ignore}[1]{}

\newcommand{\paren}[1]{\left(#1\right)}

\newcommand{\aaparen}[1]{\llangle#1\rrangle}

\newcommand{\foralle}[3]{\left(\forall #1: #2: #3\right)}

\newcommand{\figDerivationSketch}{
\begin{figure}[!t]
\centering
\includegraphics[scale=1]{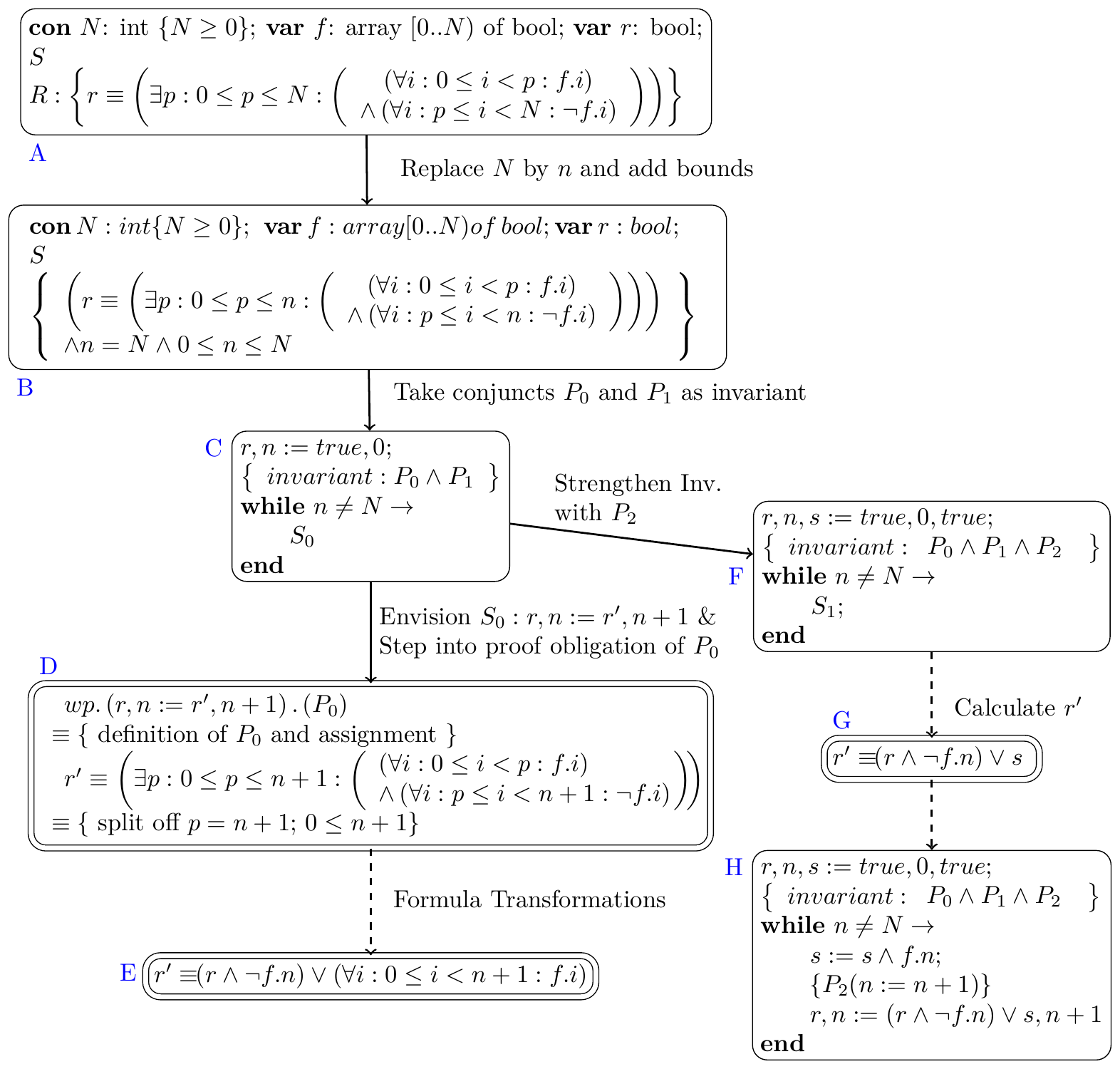}
\caption{\small Sketch of the calculational derivation for a simple program. Symbols $S$, $S_0$, and $S_1$ are the placeholders for the unknown program fragments. The single bordered boxes represent program nodes whereas the double bordered boxes represent formula nodes. \\
$P_{0}:\sp\sp\left(r\equiv\sp\left(\exists p:\sp0\le p\le n:\sp\left(
\left(\forall i:\sp0\le i<p:\sp f[i]\right)
\wedge\left(\forall i:\sp p\le i<n:\neg f[i]\right)
\right)\right)\right)$\\
$P_{1}:\sp\sp0\le n\le N; \qquad P_{2}:s\equiv\left(\forall i:\sp0\le i<n:\sp f[i]\right)$
}
\label{fig:DerivationSketch}
\end{figure}
}

\newcommand{\figNavigatingTree} {
\begin{figure}[h]
\includegraphics[scale=0.5]{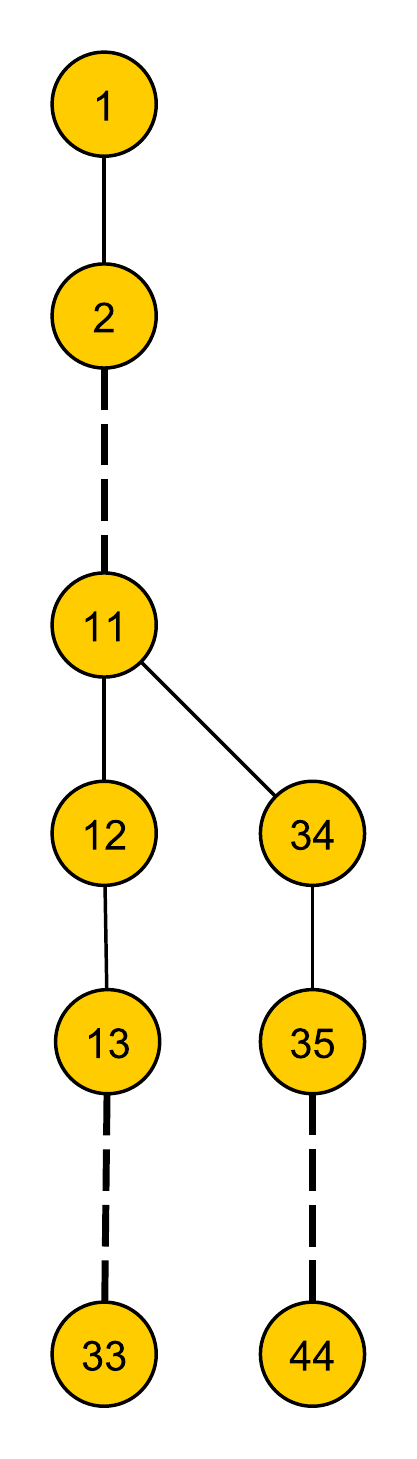}
\hfill{}
\includegraphics[scale=0.6]{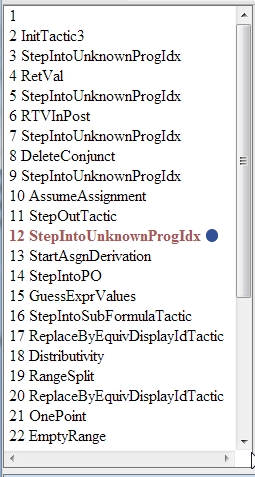}
\hfill{}
\includegraphics[scale=0.6]{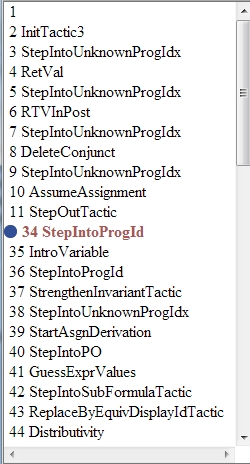}
 \caption{
 Navigating the derivation tree: Fig.~(a) shows schematic diagram of a derivation tree. Fig. (b) shows the path in the derivation tree containing the currently selected node (node 12). A marker (a filled circle) to the right of node 12 indicates the presence of a right-sibling node (node 37) in the derivation tree. Users can click on this sibling marker to switch to the branch containing \emph{node 37.}. The resulting path is shown in\emph{ }Fig. (c).
}
\label{fig:NavigatingTree}
\end{figure}
}

\newcommand{\figMainGUI}{
\begin{figure}[t]
\centering
\includegraphics[width=1.0\textwidth]{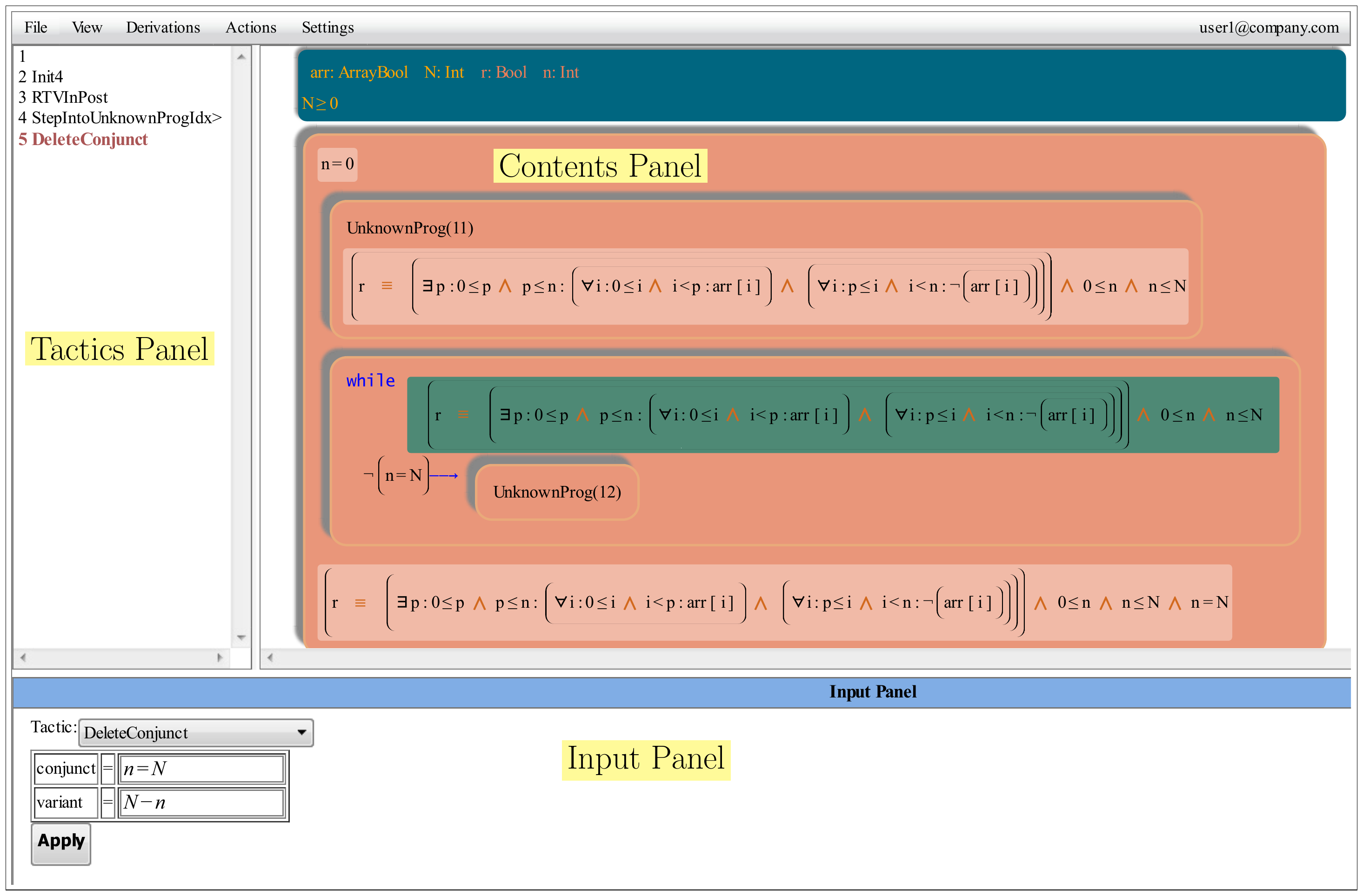}
\caption{ %
CAPS GUI}
\label{fig:MainGUI}
\end{figure}
}

\newcommand{\figStructuredRepFormula}{
\begin{figure}[t]
\centering
\includegraphics[scale=0.5]{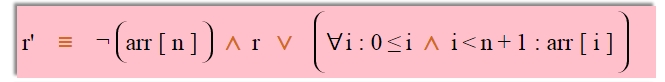}
\\
\includegraphics[scale=.5]{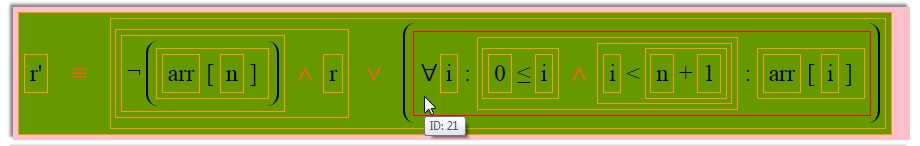} 
\caption{ %
Structured representation of a formula in \emph{normal mode} and \emph{selection mode}. Users can select a subformula by simply clicking on it.}
\label{fig:StructuredRepFormula}
\end{figure}
}

\newcommand{\figInputPanel}{
\begin{figure}[t]
\centering
\includegraphics[width=1.0\textwidth]{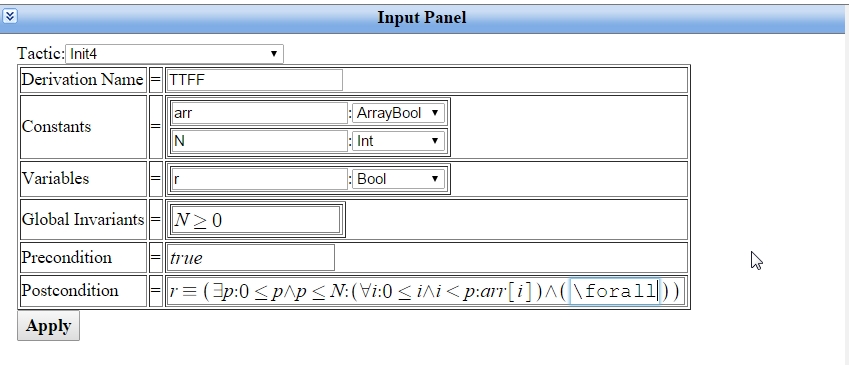}
\caption{ %
Input Panel: On selection of a tactic to be applied, the corresponding input form is dynamically generated.}
\label{fig:InputPanel}
\end{figure}
}

\newcommand{\figFormulaStructDeriv}{
\begin{figure}[t]
\centering
\includegraphics[width=0.8\textwidth]{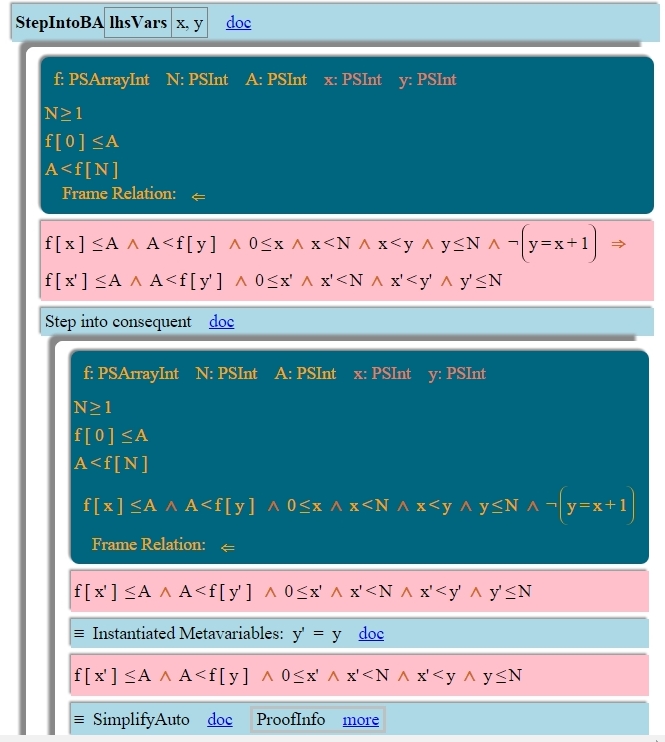}
\caption{ %
Formula transformations from the derivation of the Binary Search program.}
\label{fig:FormulaStructDeriv}
\end{figure}
}

\newcommand{\figSelDisplay}{
\begin{figure}[t]
\centering
\includegraphics[width=1.0\textwidth]{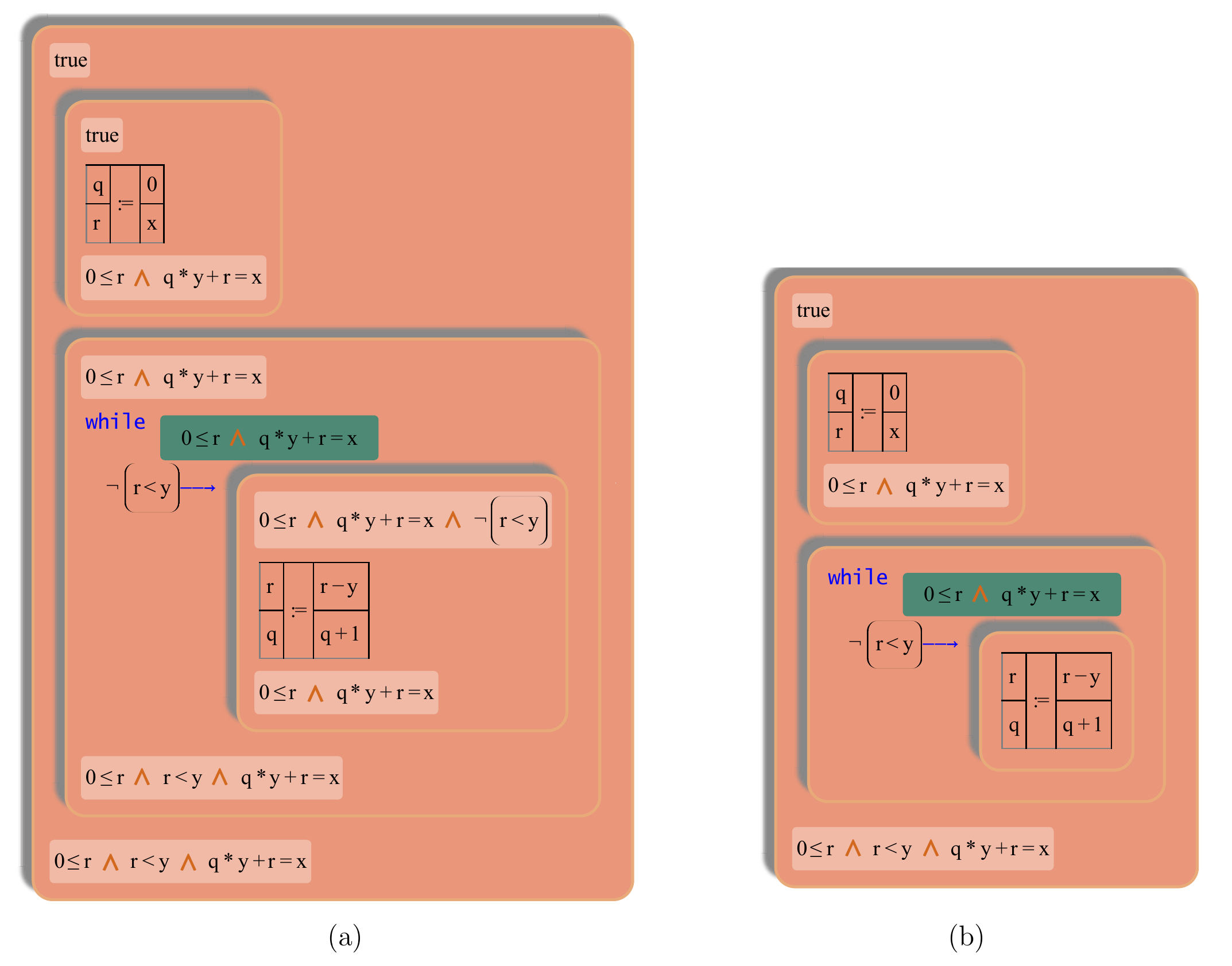}
\caption{ %
Final \emph{AnnotatedProgram} for the Integer Division problem: a) Full annotations mode, b) Minimal annotation mode.}
\label{fig:SelDisplay}
\end{figure}
}

\newcommand{\figCAPSArch}{
\begin{figure}[t]
\centering
\includegraphics[width=1.0\textwidth]{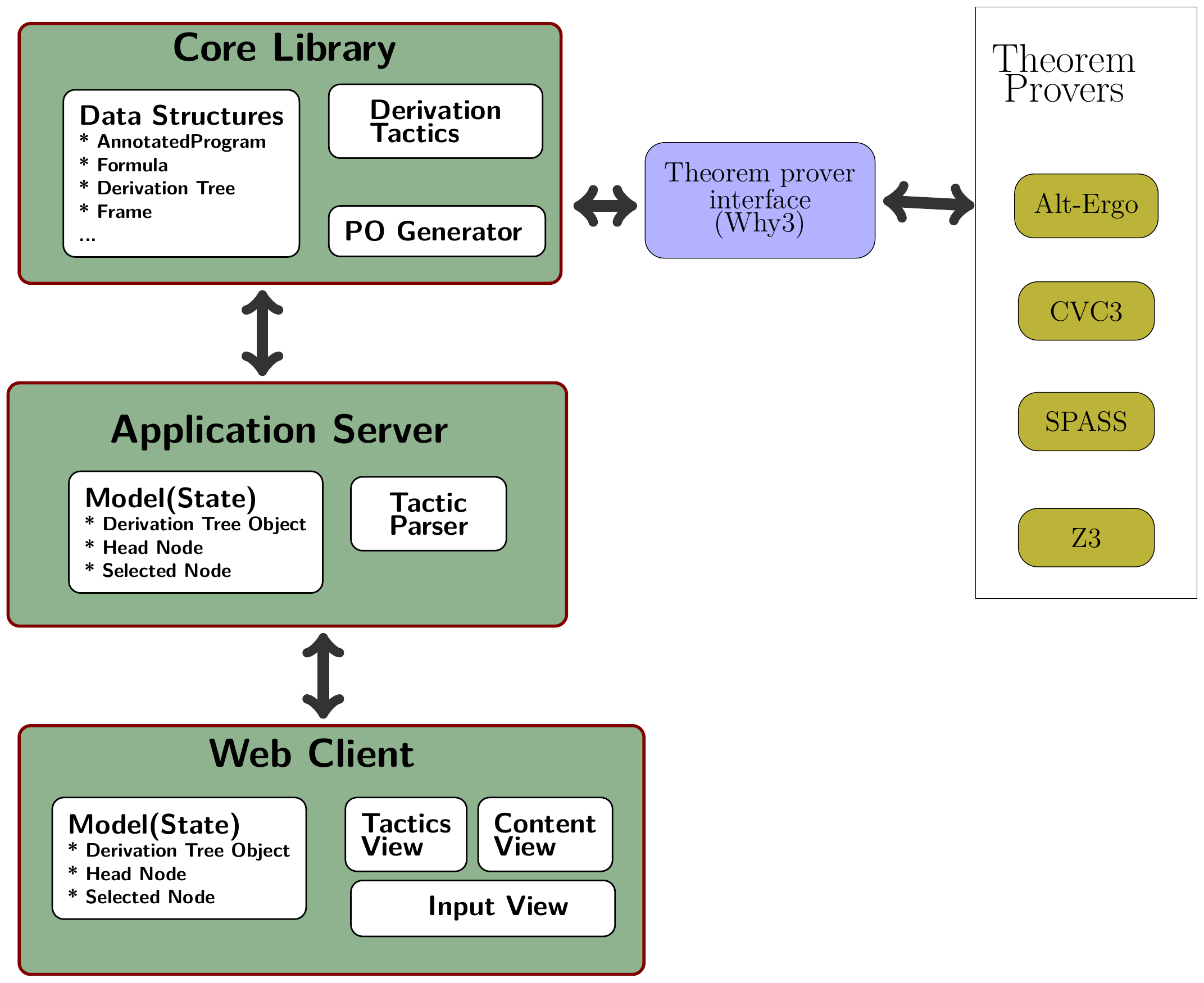}
\caption{ %
CAPS Architecture}
\label{fig:CAPSArch}
\end{figure}
}

\title{Building an IDE for the Calculational Derivation of Imperative Programs}
\author{
Dipak L. Chaudhari \qquad\qquad Om Damani
\institute{	Indian Institute of Technology Bombay, India}
\email{\quad dipakc@cse.iitb.ac.in \quad\qquad damani@cse.iitb.ac.in}
}
\def\titlerunning{Building an IDE for the Calculational Derivation of Imperative Programs}
\def\authorrunning{D. L. Chaudhari and O. Damani}

\begin{document}
\maketitle

\newcommand{\apre}[1]{\aaparen{\bm{#1}}}
\renewcommand{\floatpagefraction}{1}

\begin{abstract}
In this paper, we describe an IDE called CAPS (Calculational Assistant for Programming from Specifications) for the interactive, calculational derivation of imperative programs. In building CAPS, our aim has been to make the IDE accessible to non-experts while retaining the overall flavor of the pen-and-paper calculational style. We discuss the overall architecture of the CAPS system, the main features of the IDE, the GUI design, and the trade-offs involved.
\end{abstract}
\section{Introduction}
\label{sec:intro}
\emph{Correct by Construction} is a programming methodology, wherein programs are derived from a given formal specification of the problem to be solved, by repeatedly applying transformation rules to partially derived programs. Within this broad framework, Dijkstra and Wim Feijen~\cite{dijMethod1988} popularized the Calculational style for deriving sequential programs, where unknown program fragments are {\em calculated} from their pre- and post- conditions. By {\em calculation}, we mean that program constructs are introduced only when 
logical manipulations show them to be sufficient for discharging the correctness proof obligations.

Despite resulting in simple and elegant programs~\cite{kalProg1990}, the Calculational Style of Program Derivation did not become popular due to the various practical difficulties that prevented wider  adoption of this methodology. Even for small programming problems, the derivations are often long and difficult to organize. As a result, the derivations, if done manually, are error-prone and cumbersome.

To address these issues, we have built an IDE called CAPS (Calculational Assistant for Programming from Specifications)\footnote{CAPS is available at \url{http://www.cse.iitb.ac.in/~dipakc/CAPS}}. CAPS has built-in refinement rules and the system generates the required correctness proof obligations. In building CAPS, our aim has been to make the IDE accessible to nonexperts while  retaining the overall flavor of the pen-and-paper style derivation.

Towards this goal, in our earlier work, we described the use of theorem prover assisted tactics \cite{ChaudhariDamaniIFM14} to automate the mundane tasks during the derivations. In this paper, we discuss the overall architecture of CAPS, the main features of the IDE, the GUI design, and the design trade-offs involved. For the automation to fit into the overall calculational methodology, we have developed several features, like stepping into subcomponents, backtracking, and metavariable support. With the help of small examples, we discuss how these features address various issues with particular emphasis on usability.

\subsubsection*{Related Work.}

The Implement-and-Verify program development methodology involves an implementation phase followed by a verification phase. Tools like Why3 \cite{filliatre:hal-00789533}, Dafny \cite{leino2010dafny}, VCC \cite{cohen2009vcc} and VeriFast \cite{VerifastJacobsPiessens08} generate the proof obligations and try to automatically discharge these proof obligations.  Although the failed proof obligations provide some hint, there is no structured help available to the users in the actual task of implementing the programs. Users often rely on ad-hoc use cases and informal reasoning to guess the program constructs.

Systems like Cocktail \cite{franssen1999cocktail}, Refine \cite{oliveira2004refine}, Refinement Calculator \cite{butlerProgram1996} and PRT \cite{carringtonTool1996} provide tool support for the refinement based formal program derivation. Cocktail offers a proof-editor for first-order logic which is partially automated by a tableau based theorem prover. However, the proof style is different from the calculational style. Refine has a plug-in called Gabriel which allows users to create tactics using a tactic language called \emph{ArcAngel}. Refine and Gabriel are not integrated with theorem provers and do not support discharging of proof obligations. In case of Refinement Calculator and PRT, the program constructs need to be encoded in the language of the underlying theorem prover. In CAPS, our goal has been to be theorem-prover agnostic, so that we can exploit the advances made in different theorem provers.

The KIDS and the Specware\cite{SmithGenProg} systems provide operations for the transformational development of programs and have been very successful in synthesizing efficient scheduling algorithms. However, these systems are targeted towards expert users. Jape\cite{bornat1997jape} is a proof calculator for interactive and step-by-step construction of proofs in natural-deduction style. Although Jape supports Hoare logic, it is mainly intended for proof construction whereas CAPS is focused on program derivation and has many tactics specific to program calculations.

\section{An Example of a Calculational Derivation}
\label{sec:MotivatingExample}
\figDerivationSketch

We now present a sketch of the calculational derivation for a simple program. Consider the following programming task (adapted from exercise $4.3.4$ in \cite{kalProg1990}. The informal derivation of this problem also appears in \cite{ChaudhariDamaniIFM14}). 

\emph{Let f[0..N) be an array of booleans where N is a natural number. Derive a program for the computation of a boolean variable r such that r is true iff all the true values in the array come before all the false values.}

Fig.~\ref{fig:DerivationSketch} depicts the derivation process for this program.
We start the derivation by providing the formal specification (node \emph{A}) of the unknown program $S$. 
We apply the \emph{Replace Constant by a Variable} \cite{kalProg1990} heuristic. In particular, we replace constant $N$ by a fresh variable $n$ and add bounds on $n$ to arrive at program \emph{B}. 
After inspecting the postcondition of program shown in node \emph{B}, we decide to apply another well known heuristic \emph{Take Conjuncts as Invariants} to arrive at a \emph{While} program (node \emph{C}) with $P_0$ and $P_1$ as loop invariants. Here, $S_0$ denotes the unknown loop body. (Derivation of the initialization of the variables $r$ and $n$ is skipped.)
To ensure loop progress, we envision an assignment $r, n:=r', n+1$ for $S_0$ where $r'$ is placeholder for the unknown expression (also called a metavariable). We then step into the proof obligation for preservation of invariant $P_0$ and try to manipulate the formula with the aim of finding a program expression for the metavariable $r'$.
After several formula transformations we arrive at a formula \emph{E} $\paren{r' \equiv \paren{r \wedge \neg f[n]} \vee \foralle{i}{0 \le i < n+1}{f[i]}}$. 
At this point, we realize that we can not represent $r'$ in terms of the program variables unless we introduce a fresh variable to maintain $\foralle{i}{0 \le i < n}{f[i]}$.
We then backtrack to program $B$, introduce a fresh variable $s$ and strengthen the invariant of the \emph{While} program with $P_2$. 
For the derivation of program $S_1$, we follow the same process as that of $S_0$ with the strengthened invariant. On this derivation attempt, we are able to calculate $r'$ with the help of the newly added invariant $P_2$. Finally we derive $s := s \wedge f.n$ to establish $P_2(n := n + 1)$.\footnote{$P_2(n := n + 1)$ represents a formula obtained by textual substitution of the free occurrences of $n$ with $n + 1$ in $P_2$} The final derived program in shown in node \emph{H}. (Note that we can further improve the program by strengthening the guard.)

As can be seen in this example, the calculational derivation involves program transformations as well as formula transformations. The derivation process is non-linear involving backtracking and branching.

\section{CAPS}
\label{sec:CAPS}

In building CAPS, our aim has been to build an easy to use IDE for the calculational derivation of imperative programs. We have tried to automate the mundane tasks while striving to keep the overall approach close to the pen-and-paper calculational style. 
All the publicly available IDEs lack in one respect or another with respect to the features important for our purpose (for example, structured calculations, integration with multiple theorem provers, backtracking and branching).

\subsection{Derivation Methodology}
We use a hierarchical representation called \emph{AnnotatedProgram} for representing a program fragment along with its specification (precondition and postcondition). The \emph{AnnotatedProgram} representation can be thought of as an extension of the Guarded Command Language (GCL) \cite{dijkstraGuarded1975} where each program construct in the GCL is augmented with its precondition and postcondition. We also introduce a new program construct \emph{UnkProg} to represent an unsynthesized program fragment.
Each subprogram in the annotated program representation has its own precondition and postcondition. As we will see in section~\ref{sec:Focusing}, this hierarchical structure is helpful when the user wants to focus on each subprogram independently.

We use the formulas in sorted first-order predicate logic for expressing the precondition and the postcondition of the programs. We use the Eindhoven notation \cite{backhouse2006exercises} for expressing the quantified formulas. In the quantified formula $(OP\, i:\, R:\, T)$, The symbol $OP$ is the quantifier version of a symmetric and associative binary operator \emph{op},
$i$ is a list of quantified variables, $R$ is the \emph{Range} -
a boolean expression typically involving the quantified variables,
and $T$ is the \emph{Term} - an expression. 

Users start a derivation by providing the formal specification of a program and then incrementally transform it into a fully derived program by applying predefined transformation rules called \emph{Derivation Tactics}. 
For example, in Fig.~\ref{fig:DerivationSketch}, the user starts the derivation by providing the postcondition $R$ (node \emph{A}). 
This program is then transformed incrementally to the final program shown in node \emph{H}. During the derivation, a user might envision a subprogram in terms of the metavariables. The next task for the user is to find a program expression for the metavariable such that the proof obligation is discharged. 
This requires formula transformations to simplify the proof obligation. The derivation thus consists of the program transformations as well as the formula transformations. These derivation modes are called the program mode and the formula mode respectively. A way of transitioning between these two modes is described in section \ref{sec:Focusing}. The derivation process ends when all the unknown programs are derived. The complete derivation history is recorded in the form of the \emph{Derivation Tree}.

The final outcome of the program derivation process is the fully annotated program along with the complete derivation tree. The \emph{AnnotatedProgram} can be easily transformed to a program in a real programming language. 

\subsection{Graphical User Interface}
\figMainGUI
Fig.~\ref{fig:MainGUI} shows the Graphical User Interface of the CAPS system. It has three panels.
The central panel, also called the contents panel, shows a partially derived program (or a formula) at the current stage of the derivation. For example, the schematic node \emph{C} in Fig.~\ref{fig:DerivationSketch} corresponds to the program in the contents panel in Fig.~\ref{fig:MainGUI}. The contents in this panel can be shown at different levels of details, as discussed in section \ref{sec:SelectiveDisplay}. 
The left panel, also called the tactics panel shows the list of the tactics applied so far. It corresponds to a path the derivation tree. For example, the tactics applied from node \emph{A} to node \emph{C} in Fig.~\ref{fig:DerivationSketch} are listed in the tactics panel in Fig.~\ref{fig:MainGUI}.
Users can navigate back to an earlier point in the derivation by clicking on the corresponding node in the left panel. 
The bottom panel is the input panel. This panel is used for selecting a tactic to be applied next and for providing the corresponding tactic parameters.

\subsection{System Architecture}
\figCAPSArch

The architecture of the CAPS system is shown in Fig.~\ref{fig:CAPSArch}. There are 3 main components of the system: 
\begin{itemize}
  \item Core Library. The Core library contains the data structures for \emph{AnnotatedPrograms}, \emph{Formula}, \emph{DerivationTree}, \emph{DerivationTactic} and \emph{Frame}. It also contains a repository of the program and the formula manipulation tactics. The Core library is integrated with various automated theorem provers (Alt-Ergo, CVC3, SPASS, Z3) via the common interface provided by the \emph{Why3} framework~\cite{filliatre:hal-00789533}. The Derivation Tree management utilities are also implemented in this library. The library is implemented in \emph{Scala} and uses the \emph{Kiama} library \cite{kiama} for rewriting.
  \item Application Server. The server component is implemented using the \emph{Scala play} web framework \cite{PlayFrameworkURL}. The server stores the current state of the derivation. The application also implements a tactic parser which parses the tactic request.
  \item Web Client. The \emph{CAPS} application is implemented as a single-page web application based on the \emph{Backbone.js} framework \cite{BackboneJSURL}. The client also maintains a state of the derivation in order to reduce server trips for navigational purpose to increase responsiveness of the application. The GUI part is implemented in the \emph{Typescript} language \cite{TypescriptURL} (which complies to Javascript). The GUI module has different views to display the current state of the derivation.
\end{itemize}




\section{Textual vs Structured Representation}
\label{sec:StructuredRep}
\figStructuredRepFormula


One important decision in developing an IDE is the choice between a textual representation and a structural one. While the tools like Dafny \cite{leino2010dafny} and Why3 \cite{filliatre:hal-00789533} use textual representations, the structural representation is more suitable for a tactic based framework like CAPS. An Annotated Program in CAPS has a hierarchical structure consisting of nested programs and formulas. By Structured representation, we mean that such hierarchical elements are identifiable in the GUI. As discussed later, this allows the user to select and focus on a subprogram or a subformula. Note that doing the same in a text based representation will require extra processing~\cite{bertot1997implementing}.

Direct editing of the Annotated Program may destroy the structure and is disallowed in CAPS; the only way to generate a program is through a tactic application. This discipline allows us to capture all the design decisions taken during the derivation. However, to allow some informality, we do have tactics to directly guess a program fragment (or the next formula). In such cases, the role of a tactic application is just to ensure - with the help of theorem provers - that the transformation is correct, and that the structure is maintained.

The contents panel in Fig.~\ref{fig:MainGUI} shows the structured representation of an annotated program.
Fig.~\ref{fig:StructuredRepFormula} shows the structured representation of a formula in the normal and the selection mode.
The binary logical operators are shown using the infix notation. Only necessary parentheses are displayed assuming the usual precedence. We put more space around the lower precedence operators (like $\equiv$) to improve readability.

%

\figInputPanel
For inputting the tactic parameters, we prefer a dynamically generated GUI instead of a static textual input form. On selecting a tactic to be applied next, the corresponding input form is dynamically generated. Users need not remember the input parameters required for the tactic.  Fig.~\ref{fig:InputPanel} shows the tactic input panel for the \emph{Init4} tactic which is used for specifying the program. Since CAPS is a web-based application, the hypertext-based display enables providing a help menu for input parameters in a user-friendly way.

For entering formulas, however, we prefer textual input. The formulas are entered in the Latex format. The formula input box is responsive; as soon as a Latex expression is typed, it converts the expression into the corresponding symbol immediately.

\section{Focusing on subcomponents}
\label{sec:Focusing}
During the program derivation process, an annotated program is nothing but a partially derived program containing multiple unsynthesized subprograms. The derivation of these unsynthesized subprograms is, for the most part, independent of the rest of the program. Hence the CAPS system provides a facility to extract all the contextual information required for the derivation of a subprogram so that the user can focus their attention on the derivation of one of these unknown subprograms. A subprogram can be selected by simply clicking on it. On selecting a subprogram, only the extracted context of the subprogram, and its precondition and postcondition are shown whereas the rest of the program is hidden.

\figFormulaStructDeriv

Similar to the subprogram extraction, users can chose to restrict attention to a subformula of the formula under consideration. On focusing on a subformula, the system extracts and presents the contextual information necessary for manipulating the subformula.

Our subformula representation is an extension of the \emph{Structured Calculational Proof} format \cite{back1997structured}. The implementation details and the theoretical basis of the contextual extraction is given in \cite{ChaudhariDamaniIFM14}.

Fig.~\ref{fig:FormulaStructDeriv} shows a snapshot of the formula transformations involved in the derivation of the binary search program. The derivation is displayed in a nested fashion. Whenever the user focuses on a subformula, an inner frame is created inside the outer frame. The assumptions available in each frame are displayed on the top of the frame. In the figure, as the user focuses on the consequent of the implication, the antecedent is added to the assumptions. On successful derivation of all the metavariables, user can step out from the formula mode to create a program where the metavariables are replaced with the corresponding derived expressions. 

Unlike the hierarchical program structure, the hierarchical formula structure is not usually shown in the GUI. This is done to reduce the clutter as the hierarchical formula structure can get very large. It is only displayed when we intend to select a subformula. This user interaction mode, called a selection mode, is used to select subformulas to be focused on. Fig.~\ref{fig:StructuredRepFormula} shows a formula in the normal mode and in the selection mode.

\section{Selective Display of Information}
\label{sec:SelectiveDisplay}
\figSelDisplay
In the \emph{AnnotatedProgram} representation, all the subprograms are annotated with the respective precondition and postcondition. Although this creates a nice hierarchical structure, it results in a cluttered display which places higher cognitive demand on the attention and mental resources of the users. An effective way to keep the cognitive load  low, is to hide information that is not relevant in any given context, such as the annotations that can be easily inferred from the other annotations.
CAPS provides a \emph{Minimal Annotations} mode which displays only the following annotations.

\begin{itemize}
\item Precondition and postcondition of the outermost program
\item Loop invariants
\item The intermediate-assertion of the \emph{Composition} construct
\end{itemize}

All other annotations can be inferred from these annotations without performing a textual substitution required for computing the weakest precondition with respect to an assignment statement.
Fig.~\ref{fig:SelDisplay} shows the Integer Division program with full annotations and with minimal annotations. All the hidden annotations can be easily inferred from the displayed annotations. The minimal annotations reduce the clutter to a great extent. 

In addition to the annotations, there are lots of other details that can be hidden. For example, the discharge status of various proof obligations for the \emph{SimplifyAuto} tactic can run into several pages, and is hidden by default (The \emph{ProofInfo} link in the Fig.~\ref{fig:FormulaStructDeriv}). The annotated programs can also be collapsed by double clicking on them.

\section{Maintaining Derivation History}
\label{sec:history}
Invariant and assertion annotations help in understanding and verifying a program. However, they provide little clue about how the program designer might have discovered them. For example, at node \emph{E} in the derivation in Fig.~\ref{fig:DerivationSketch}, we are unable to express the expression under consideration in terms of the program variables. This guides us to introduce a fresh variable $s$ and strengthen the invariant with $P_{2}$. This crucial information is missing from the final annotated program. It is therefore desirable to preserve the complete derivation history to fully understand the derivation of the program. CAPS maintains the derivation history in the form a derivation tree. Maintaining history also facilitates backtracking and branching if the user wants to try out an alternative derivation strategy.

\subsection*{Backtracking and Branching.}
In CAPS, we do not allow programmers to directly edit the program; users have to backtrack and branch to try out different derivation strategies. This restriction ensures that the derivation tree contains all the information  necessary to reconstruct the program from scratch. All the design decisions are manifest in the derivation tree which helps in understanding the rationale behind the introduction of various program constructs and invariants. Using the branching functionality, users can explore multiple solutions for the given programming task.


%
%


\subsection*{Navigating the Derivation tree }

\figNavigatingTree

The conventional tree interface is not suitable to showing the derivation tree. At any point during a derivation, we are interested in only the active path of the derivation tree. This active path is shown in the left panel in the GUI. To make it easy to navigate to other branches, we show siblings of the nodes in the path. Users can navigate across the branches by clicking the sibling markers as shown in Fig.~\ref{fig:NavigatingTree}. If there are multiple branches under the selected sibling, then the rightmost branch is selected.

\section{Conclusions and Future Work} 
In this work, we have described the design of an IDE for the Calculational Derivation of Imperative Programs. Our design focus has been on making the IDE accessible to nonexperts while retaining the overall flavor of the pen-and-paper style derivation.  We have used the CAPS system in an elective course on \emph{Program Derivation} taken by 2nd year students. The preliminary student response to the
tool has been very positive\cite{ChaudhariDamaniITICSE15}. However, a thorough evaluation needs to be done on more challenging problems.

Based on the learnings from the first offering of the tool, we plan to enhance the tool in a number of ways.

\subsubsection*{Richer Language Constructs.}
We plan to target programs with richer constructs involving recursion, algebraic data types, and polymorphic types.

\subsubsection*{Executing programs. } We currently do not have a functionality to execute the derived programs in CAPS. We plan to explore the possibility of executing not only the final program, but also the intermediate partially derived programs. Being able to simulate programs at the intermediate stages of behavioral abstraction has already been identified\cite{LeinoDirection} as one of the barriers in the adoption of the stepwise refinement based methods.

\subsubsection*{Integrating Synthesis Solvers.} We plan employ the synthesis solvers \cite{alur-fmcad13} during the interactive derivation when the specification of the subprogram under consideration falls in a theory for which a synthesis solver is available. We will, however, restrict the use of these solvers to the synthesis of loop-free programs. 

\subsection*{Acknowledgements.}
The work of the first author was supported by the Tata Consultancy Services (TCS) Research Fellowship and a grant from the Ministry of Human Resource Development, Government of India.


\bibliographystyle{eptcs}
\bibliography{FIDE15}
\end{document}